# Raman signal-enhancement and broadening with graphene-coated diamond-shape nano-antennas

*Charilaos Paraskevaidis[1], Tevye Kuykendall[2], Mauro Melli[2], Alexander Weber-Bargioni[2], P. James Schuck[2], Adam Schwartzberg[2], Scott Dhuey[2], Stefano Cabrini[2] and Haim Grebel[1\*]*

[1] *Electronic Imaging Center, NJIT, Newark, NJ 07102;* <grebel@njit.edu>

[2] *The Molecular Foundry, Lawrence Berkeley National Laboratory, Berkeley, CA 94720*

**Abstract:** We used broad-band diamond-shape antennas, whose bandwidth could cover the frequency range between the pump laser and the scattered modes, in conjunction with uniformly deposited graphene test films. We demonstrated amplification by two orders of magnitude of the Raman D-line, accompanied by 5-fold line broadening, never seen before. The line broadening decreased as a function of increasing line peak frequency, as if mapped by the antenna's gain curve. Understanding line broadening observed with broad-band nano-antennas is key to proper interpretation of molecular vibrations in SERS applications. On the other hand, other applications may be envisioned: broadband nonlinear scatterings are inherent part of optical nano-amplifiers.

Surface enhanced Raman scattering (SERS) [1] is a major spectroscopic tool for an efficient molecular identification down to the level of a few molecules [2,3]. Driven by the notion of localized electric field intensities, research turned to aggregates of metal particles [4] and then to the more structured nano-optical antennas [5-7]. Yet, most SERS antennas are relatively narrow band; covering mostly the laser frequency. Antennas covering only the scattered mode (the Raman line) deemed to be inefficient [8]. Antennas in the form of tip (tip enhanced Raman scattering, TERS) [9] couple the near-field molecular vibrations to current monopoles along the tip. Yet, lack of directivity, formation of travelling waves between substrate and antenna and formation of many radiation modes at the tip gap, limit the coherence of the frequency band to mainly the laser frequency. A trait of narrow band optical antennas is the amplification of mainly the incident local fields, amplifying equally all Raman lines. Enhancement of the Raman signal has been further hindered by uncertainties in analyte concentration and location relative to the antenna geometry.

Most practical systems assess Raman scattering at the far-field, or many wavelengths away from the sample. Therefore, localized vibrations need to efficiently couple to radiative antenna modes in order to be properly detected; the antenna should absorb the pump frequency and radiate the scattering frequency. Maintaining coherence between the pump and scattered modes is key to an efficient process. Here we examined diamond-shape antennas (D-ant) [10]. Its bandwidth accommodates the frequency range between the pump laser and the scattered Raman frequencies and it exhibits linear phase throughout that bandwidth. Effectively, the antenna increases the molecular cross-section for the radiated Raman scattering. By using graphene [11] as a test film we eliminated much of the uncertainty in molecule's location and concentration, thus enabling better comparative studies.  We not only observed large signal amplification but also large line broadening (e.g., 5-fold for the graphene D-line), unseen before.  Ramification of these findings goes beyond SERS experiments; we may envision applications in optical amplifiers as well.  As a general comment; one should distinguish between coherent broadband antennas (capable of generating short pulses, and therefore exhibiting linear phase throughout their gain bandwidth) and incoherent broadband antennas (capable of transmitting a large bands of frequencies but do not exhibit linear phase throughout their gain bandwidth; for example, the log periodic antenna at microwave frequencies).  Many antennas may accommodate the fundamental frequency and its harmonics but that does not turn them broad band.

Figure 1a,b show typical near-field simulations of scattering and near-field intensity distribution for D-ant and bow-tie antennas (BT-ant) at resonance.  As an example, the following antenna dimensions were used: height 74.1 nm and base 81.7 nm.  The spectral bandwidth of the D-ant is ca 50% wider than the BT-ant.  The peak scattering shifts up as the antenna's height becomes larger and shifts down as the gap becomes larger.  Accepted dispersion relations for graphene and gold were used [12,13].  Unlike BT-ant, 'hot spots' appear here at the antenna's extremities.  Figure 1c shows simulations of the antenna's far-field spectral scattering for rounded tip antennas with a gap of 12 nm.  The simulations underscore the bandwidth advantage of diamond over bowtie antennas.  While the nominal dimensions of the D-ant and BT-ant were the same, the fabrication process unintentionally may have increased the gap of BT-ant by rounding their apices. Such is the case shown in Fig. 1c: the rounding of the antenna's tip increased the gap from 12 nm (for sharp tips) to 22 nm (with round tips).

Figure 2a is a scanning electron microscope (SEM) image of our layout. Each antenna was surrounded by fiduciary marks, 2 μm away from the antenna. The rectangular patterns were oriented perpendicularly to one another in order to examine the effect of two linearly orthogonal incident polarizations states. The entire metal layout exhibited photoluminescence (PL) when illuminated by the 633 nm HeNe laser [14] and as shown by Figure 2c. The PL signal of gold was small when illuminated with the 785 nm laser and thus, mostly the graphene lines on antennas were accentuated near resonance (Fig. 2d). The background reflection has substantially elevated near resonance: this occurred because of ordinary antenna reflectivity (it is made of metal after all), because the antenna at resonance exhibited directivity gain (namely, scatters more light at normal direction to its axis), and because gold luminesces above ~550 nm. Such scattering necessitate background correction. The background was fitted by a polynomial and subtracted from the data using two different routines to achieve similar results. In the SI section we show examples of uncorrected Raman data for BT and D antennas (Fig S7). The entire data is limited at the lower end by a cut-off laser filter and a small background attributed to the oxide substrate (0-500 $cm^{-1}$). The upper end is limited by the CCD dark current. In most cases we fitted the curves in the frequency range where three major garaphene lines may be identified; 1300, 1600 and 2700 $cm^{-1}$ (or the 1000-3000 $cm^{-1}$ range). The amplification and broadening of the small 800 $cm^{-1}$ line (attributed to pure amorphous carbon [15] present in the film) became so large that it overwhelmed the spectral region between 500-1000 $cm^{-1}$.

Background corrected images are shown in Figure 3 (see also the supplementary information (SI) section). We did not observe the two 'hot-spot' regions at the antenna's extremity despite a better than 120 nm scan accuracy in some cases; either because the close-proximity 'hot spot' regions could not be resolved in the far-field, or, that the notion of 'hot spot' does not apply to our case. When the polarization of the pump laser was made perpendicular to the antenna's axis, scatterings from one subset of the fiduciary marks became more pronounced. Typically, the narrow side of the fiduciary marks (ca 100 nm ) and the disks exhibited data similar to the BT-ant.

Each of the antenna's spatial peak (e.g., of Figure 3) was fitted by a Gaussian distribution. The average spatial width at Raman frequencies (full width at half maximum, FWHM) was assessed as 1.06±0.02 $\mu$m (this fitting error should not to be confused with the scan resolution). This value is 1.4 times larger than the FWHM of the laser spot-size (assessed at 0.715 $\mu$m by a Gaussian beam approximation). What it means is that, at the Raman frequencies, the antenna's cross section became larger than the diffraction limited spot, and is typical of resonance conditions (care was taken that the laser beam was in focus before and after scans; this conclusion is based on observation from several scans in various antenna regions). By comparison, the scan over the (rated) 10-nm gap BT-ant exhibited an antenna width of approximately 0.8 micron wide only slightly larger than the laser spot size. The scan of D-ant with perpendicular polarization exhibited effective antenna width of 0.9±0.02 microns, alluding to its only partial resonance.

In Figure 4 we compare the spectra of graphene with the laser focused on top of the D-ant antenna and in between antennas. Large peak enhancement (by comparing peaks amplitudes, or counts) on the order of 100 times and a 5-fold peak broadening of the 1300 cm$^{-1}$ line (D-line) are characteristics of D-ant at resonance. Enhancement and broadening of the 1600 cm$^{-1}$ line (G-line) was typically smaller: 10-fold enhancement and 3-fold broadening. The 2700 cm$^{-1}$ line was broadened by a factor of 1.5 with little enhancement. The largest identifiable enhancement and line broadening was observed for the 1340 cm$^{-1}$ graphene line - closer to the resonance peak. The smaller enhancement and broadening was detected for the 2700 cm$^{-1}$ graphene line – furthest from the resonance peak. As if, the antenna gain curve was mapped onto the Raman lines and superimposed onto the detector responsivity. The 2D line is of similar strength to the G-line when measuring in between antennas or using the 633 nm laser (see SI section). To a lesser extent, BT-ant exhibited similar enhancement and broadening trend. For example, enhancement values were 6 and 5 with broadening of 3 and 2 times for the D and G line respectively for the rated 10 nm BT-ant. Polarization dependent enhancement of Raman lines for graphene over the fiduciary marks was also observed and varied in intensity. The 100 nm narrow size of the rectangular shape mark and the 100 nm disk shape diameter were within resonance range for the 785 nm laser. Overall, the spectra for graphene-coated D-ant shifted down a bit (by 2% for the background corrected data) while there was no, or very little shift for BT-ant and fiduciary marks.

Is line broadening the result of amorphous carbon at the antenna's 'hot spots'? The argument goes as follows: amorphous carbon [16, 17] formed at the antenna's tip by photo-carbonation at the hot-spots areas and amplified the D and G-lines of graphene on the antennas. Accordingly, these lines would be uncharacteristically enhanced because of the 'hot spot' contribution to the overall signal. Against this argument we point to the following: (a) the broadening of the 2D line (Figure 4) cannot be easily explained by this argument since the overall signal should have been dominated by the burnt graphene spots. (b) BT-ant exhibit more intense 'hot spots' even when a bit off-resonance. Why then the BT-ant exhibited smaller enhancement and smaller line broadening than the D-ant? (c) Plots shown in Fig. S4c. (laser polarization is perpendicular to the antenna axis) were taken immediately after the scan of Figure 5d (laser polarization parallel to the antenna axis). While the signal amplification in Fig. S4c. is obviously smaller, the line broadening should have been the same if the tip has burnt in the previous scans. In contrast, the D- and G-line widths of Fig. S4c are smaller by an average of at least 10% for both lines than the corresponding line widths of Figure 5d (we fitted each peak in the scans by two Gaussians to be convinced). This is easily explained if amplification and broadening are related to the antenna resonance and not to the hot spots. (d) No sign of broadening observed in data taken with the 633 nm laser from the same antennas used after the conclusion of the experiments with the 785 nm laser (sample D_30_09 of the SI section). (e) Line splitting and down-shifting for a graphene flake depositing high pillar antennas [8] were previously assigned to a linear stress; however, one may well argue that the linear stress incurred should also be exhibited in the spontaneous Raman data [18,19] regardless of whether the pump or the probe frequencies are at resonance with the antenna. (f) Data accumulation of 2 sec resulted in twice the count obtained with 1 sec while exhibiting the same D-line broadening and suggesting no lingering damage. For all of these reasons we conclude that the enhancement and line broadening are the result of collective scatterings by the entire antenna.

Based on simulations (Fig. 1c) we anticipated a large difference between similarly rated D- and BT-ant. In Figure 5a,b we present uncorrected maps of the 2700 cm$^{-1}$ graphene line for the rated 10-nm gap antennas. The D-ants are clearly visible (only two rows have been scanned) whereas the signals from fiduciary marks are much dimmer. Figure 5c shows uncorrected spectra for graphene on each antenna; the uncorrected curves include an elevated background near resonance. Besides an obvious enhancement and substantial broadening of the 1340 cm$^{-1}$ and the 1600 cm$^{-1}$ lines, D-ants exhibited bandpass characteristics between the cut-off filter and ca 1700 cm$^{-1}$. A 800 cm$^{-1}$ line may be observed 'riding' on that background, as well. In contrast, the BT-ant exhibited a monotonic tail away from the laser line. The height of this tail varied, yet, an increased tail height did not lend itself to larger Raman signals and sometimes to the contrary. The monotonous tail was also characteristic of fiduciary marks and could be used to distinguish resonating and partly resonating structures.

Why line broadening has not been seen before (excluding intentional stress related effects)? We point out that our pump wavelength is beyond the effective PL range of gold, minimizing much of the background noise. We also point out that when the entire spectrum is examined, a better picture is obtained. Our antennas have relatively large gain and are broad enough to enable coherent scattering effects. Gain has funneled to specific lines, with all graphene atoms on the antenna radiating collectively. If our model is correct, the largest amplification and broadening would be seen for the 800 cm$^{-1}$ line because it is the closest to the antenna's peak resonance. If true, this line has been broadened beyond recognition. Selective line broadening was observed for the D- and BT- antennas (see SI section), namely, only the D-line was broadened but not the others. These data could be explained by a shorter, antenna's peak wavelength resonance and in the case of the BT-ant, narrower bandwidth. Coherent plasmonic effects coupled with graphene have been observed with resonating structures before [20,21].

White light reflection experiments were carried out in order to determine the peak reflectance of the various antennas and are shown in the SI section. These were found useful in assessing peak wavelength reflection for a given set of antennas.

*On signal amplification by resonating elements:* (1) Coupling of local molecular vibrations to electromagnetic radiation at the far-field requires an efficient antenna. The graphene and all other layers) should be incorporated into the scattering simulations (as done here). (2) Excitation of the surface plasmons requires optical resonance at the incident and scattered frequencies [22]. Our data suggest that the gain was funneled to specific vibrations lines. Raman maps are, typically, spatially limited by the laser spot-size. However, the ca 100 nm D-ant was resolved to larger than the laser spot size at resonating wavelengths (the 785 nm laser). This means that the scattering cross section for the D- or G-lines has increased by a factor larger than 3 at resonance; this was due to the large gain (directivity) of our antennas. (3) A qualitative model presented in the SI section suggests that two major factors in signal enhancement are the antenna's center resonance wavelength and its resonance bandwidth. (4) It is interesting to compare the overall local field effect for BT and D-ant. We use Figure 1 as a guide. Let us assume that the antenna's peak resonance is situated between the pump and scattered lines and both lines are contained within the antenna's bandwidth. Along with the hot-spots model the Raman signal enhancement is proportional to $N|E_L|^2|E_S|^2$ [20]. Here: $E_L$ is the local field of the laser line; $E_S$ is the local field of the scattered (Raman) mode; for spontaneous Raman process $|E_S|^2 \sim \sigma|E_L|^2$ where $\sigma$ is the scattering cross section; N is the number of graphene atoms involved and is proportional to the graphene area. Note that the graphene covers the entire antenna surface. While the local field within the gap of the D-ant is smaller than the local field within the gap of BT-ant, its area is larger. Also, each apex of the antenna contributes to the signal enhancement, as well. It can be easily shown that under these terms, the enhancement advantage of the D-ant over the BT-ant would be only 10%. The experimental data suggest a much larger advantage of the D-ant over BT-ant. While broadening under local field amplification (e.g., via Rabi flopping frequency) cannot be rules out, these have been observed with specific combination of molecule/antenna hybrids [22].

In summary, we demonstrated selective Raman signal enhancement and line broadening with graphene-coated diamond shape nano-antennas. Such results may open the door for nano-optical amplifiers.

**Methods**

**Antennas layout and fabrication**. Ti/Au (3 nm/17 nm) nano-antennas were fabricated on 4" quartz wafers using e-beam lithography. These were of diamond shape (D-ant) and bowtie (BT-ant), with three gap values of 10, 20 and two 30 nm for the D-ant and 5, 10 and two 20 nm for the BT-ant. Approximately 500 antennas were fabricated for a given gap value and dose; nine e-doses were used; in total there were ~4500 antennas per given gap value, albeit variation in antennas efficiency is noted (see SI section Figs. S3, S4). It was initially expected that the antennas would not be easily identified so we included fiduciary marks to the antenna layout. These were placed 2 μm away from each antenna. The marks were either rectangular (100 x 300 nm) or disk (100 nm in diameter) shapes and were made of Ti/Au (3nm/17 nm). The upper gold surface was coated with 2 nm thick amorphous $Al_2O_3$ film using atomic layer deposition (ALD), thus separating the conductive graphene from the antenna; in this way we eliminated the chemical enhancement contribution to the Raman signal, as well.

**Graphene CVD deposition.** Graphene layers were grown in an Aixtron "Black Magic" system on 4", 25 μm copper foils (Sigma-Aldrich) and were transferred onto the antenna layout. The transfer was completed by first depositing a layer of 300 nm thick poly(methyl methacrylate) (PMMA) and subsequently etching the copper foil by $FeCl_3$ [22]. The PMMA film was later removed by immersing the samples in acetone; however, remnants of polymer were observed in the scanning electron microscope images.

**Spatial and spectral scans.** NT-MDT Raman system in confocal mode was used with a 30 mW 633 nm HeNe gas laser and a 70 mW 785 nm semiconductor laser. The laser intensity at the sample has decreased by more than 10 dB of that value due to filtering and scattering. The system was equipped with a translational stage which maintained 2-nm accuracy for over 2 hours. The laser beam was focused before every scan and its focusing after the scan was ascertained. The spectrometer was equipped with a cooled Si CCD array (Andor). Spectrometer gratings were 600 grooves/mm (blazed at 600 nm) when using the 633 nm laser and 150 grooves/mm (blazed at 500 nm) when using the 785 nm laser. The latter grating accommodated the wide spectral range of Raman scattering associated with the 785 nm laser and its line width of 0.5 nm. We note that the broadening of the Raman lines were real: it exceeded the laser line width by far. In terms of wavelengths, the 2700 cm$^{-1}$ graphene line was situated at $\lambda$=985 nm, fairly close to the camera's sensitivity edge. We used an achromatic 100x objective lens (NA 0.7) to focus the laser beam onto the sample. Data accumulation times for each point were 0.5 s for the 633 nm laser and 1 and 2 s for the 785 nm laser. SEM pictures taken after laser scanning ascertained that the antennas were not damaged during the experiments (e.g., Fig. 2b).

**Simulations and spectral analysis.** Commercial COMSOL code was used to simulate the various structures. The simulated full structure consisted of a quartz substrate, metal, surrounded by 2 nm $Al_2O_3$, two layers of suspended graphene, topped by 25 nm PMMA. Background corrections were made by using two routines: one was provided by the NT-MDT system software and the other was ours, based on a MathWorks background correcting module. Both codes gave similar results regarding the peaks value and its width. Finally, we present the data without smoothing – it might be less pleasant to the eye but provide the reader with the actual data; e.g., the peak at ~1470 cm$^{-1}$ is real (Fig. 4).

**Figure Captions**

**Figure 1**: **Electric field intensity distribution and scattering pattern**. Near-field distribution for (**a**) D-ant and (**b**) for BT-ant. (**c**) Normalized spectral reflection distribution. The two solid lines are for rounded tips at a 12 nm gap. Simulations when rounding of a nominally 10 nm gap BT-ant resulted in a 22 nm gap are represented by the dash curve.

**Figure 2**: **Electronic and optical scattering images.** (**a**) SEM image of the layout: the yellow arrows point to the antennas; the light blue arrows point to the marks. (**b**) Graphene coated BT-ant (top) and D-ant (bottom). (**c**) Uncorrected image of the layout as viewed at λ=704 nm (the 1600 $cm^{-1}$ line of graphene when the sample was illuminated by a 633 nm HeNe laser). The figure is rotated by 90 degrees when compared with the layout in (**a**). (**d**) Uncorrected image of the layout as viewed at λ=895.5 nm (the 1600 $cm^{-1}$ line of graphene when the sample was illuminated with a 785 nm laser)

**Figure 3**: **Background corrected maps.** (**a**) At λ=873.5 nm (the 1340 $cm^{-1}$ line of graphene, illuminated with λ=785 nm laser). (**b**) At λ=895.5 nm (the 1600 $cm^{-1}$ line of graphene, illuminated with λ=785 nm laser). Both **a,** and **b,** were taken with laser polarization parallel to the antenna's axis. (**c**) At λ=873.5 nm, yet with polarization perpendicular to the antenna's axis. A slight shift of the stage is noted. The antenna's gap of D-ant was nominally 10 nm.

**Figure 4**: **Raman spectra.** Graphene-coated, nominally rated 10-nm gap D-ant: Raman spectrum obtained between antennas and at the antenna's center.

**Figure 5**: **Spatial maps and Raman spectra. (a**) Uncorrected maps of BT-ant at 2700 cm$^{-1}$. (**b**) Uncorrected maps of D-ant at 2700 cm$^{-1}$. The laser wavelength was 785 nm and the antenna's gap was nominally 10 nm. The yellow arrows point to the antennas and the light blue arrows point to the marks. An optical filter cut the laser line near zero for both antennas. (**c**) Uncorrected Raman spectra of both antenna type. (**d**) Background corrected Raman spectra. The inset shows a typical spectrum of graphene between the D-ant; the relative small amplitude of the 2700 cm$^{-1}$ is due to the limited sensitivity of the Si detector array in this spectral range (~980 nm).

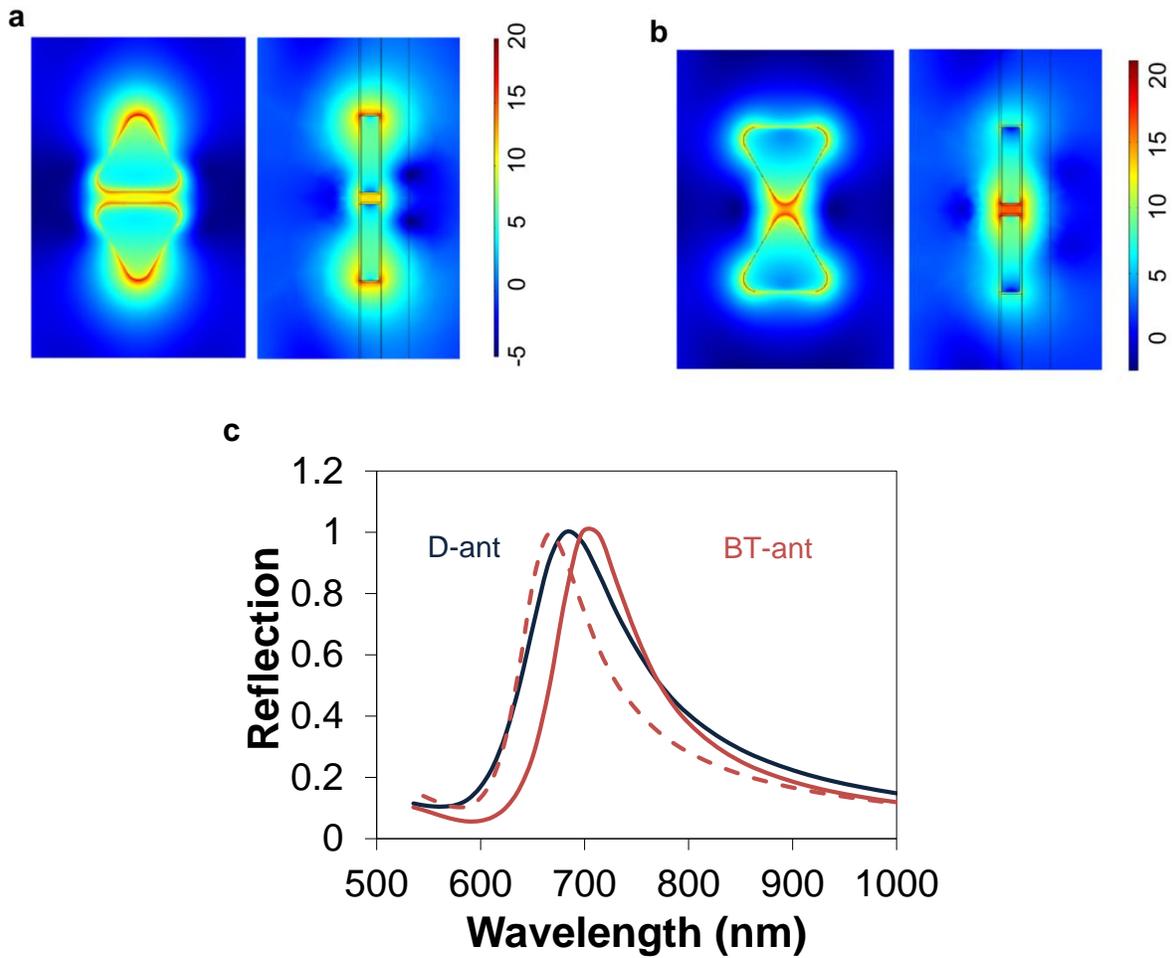

**Figure 1: Electric field intensity distribution and scattering pattern**. Near-field distribution for (**a**) D-ant and (**b**) for BT-ant. (**c**) Normalized spectral reflection distribution. The two solid lines are for rounded tips at a 12 nm gap. Simulations when rounding of a nominally 10 nm gap BT-ant resulted in a 22 nm gap are represented by the dash curve.

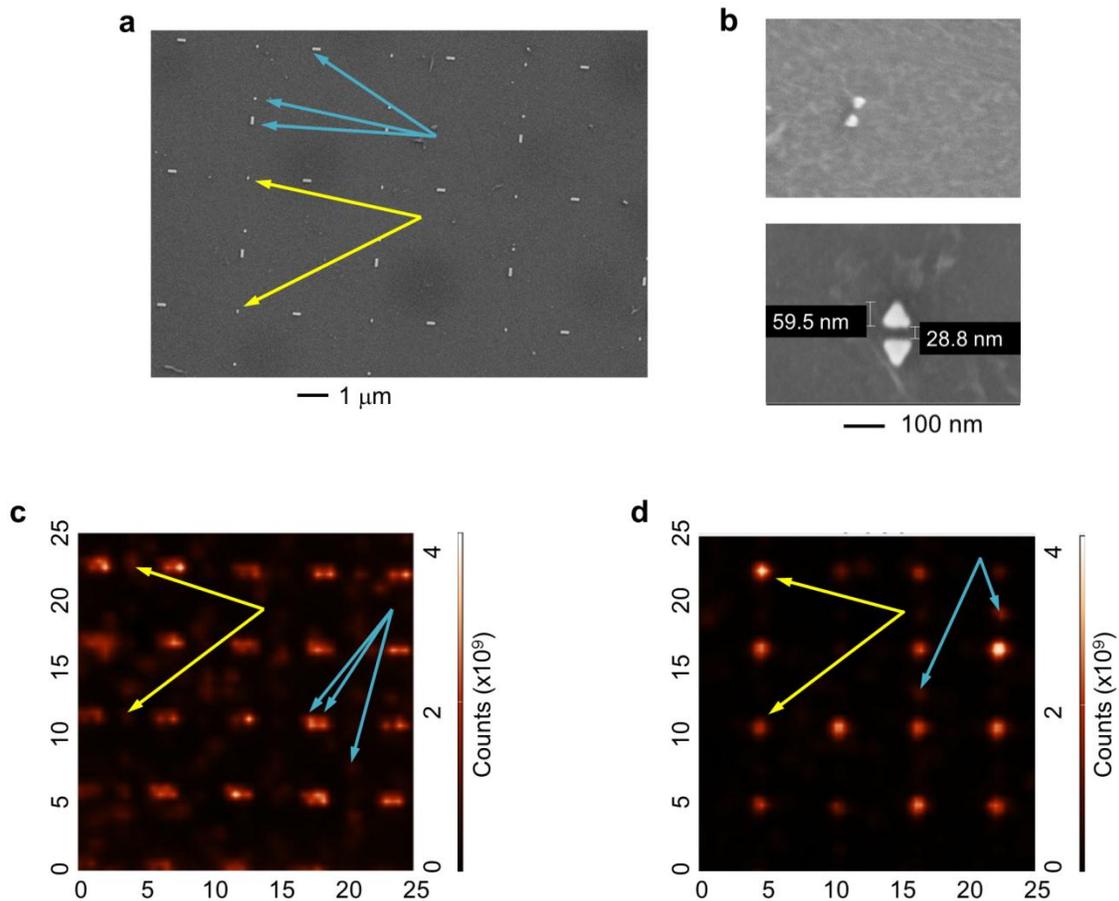

**Figure 2**: **Electronic and optical scattering images.** (**a**) SEM image of the layout: the yellow arrows point to the antennas; the light blue arrows point to the marks. (**b**) Graphene coated BT-ant (top) and D-ant (bottom). (**c**) Uncorrected image of the layout as viewed at λ=704 nm (the 1600 cm$^{-1}$ line of graphene when the sample was illuminated by a 633 nm HeNe laser). The figure is rotated by 90 degrees when compared with the layout in (**a**). (**d**) Uncorrected image of the layout as viewed at λ=895.5 nm (the 1600 cm$^{-1}$ line of graphene when the sample was illuminated with a 785 nm laser)

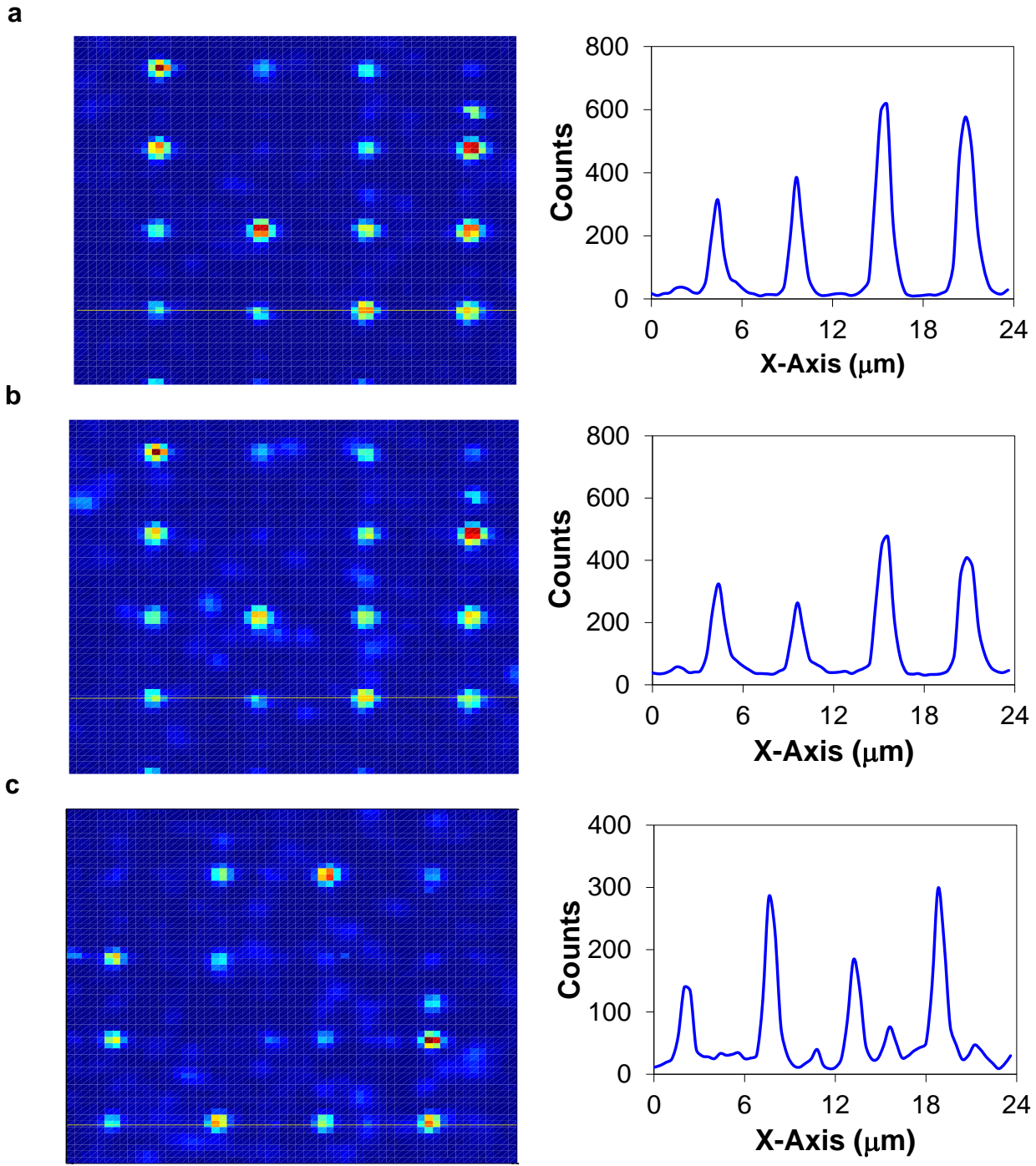

**Figure 3: Background corrected maps.** (**a**) At λ=873.5 nm (the 1340 cm$^{-1}$ line of graphene, illuminated with λ=785 nm laser). (**b**) At λ=895.5 nm (the 1600 cm$^{-1}$ line of graphene, illuminated with λ=785 nm laser). Both **a,** and **b,** were taken with laser polarization parallel to the antenna's axis. (**c**) At λ=873.5 nm, yet with polarization perpendicular to the antenna's axis. A slight shift of the stage is noted. The antenna's gap of D-ant was nominally 10 nm..

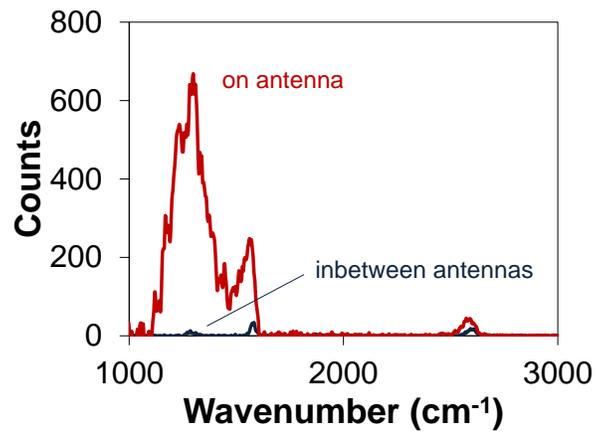

**Figure 4**: **Raman spectra.** Graphene-coated, nominally rated 10-nm gap D-ant: Raman spectrum obtained between antennas and at the antenna's center.

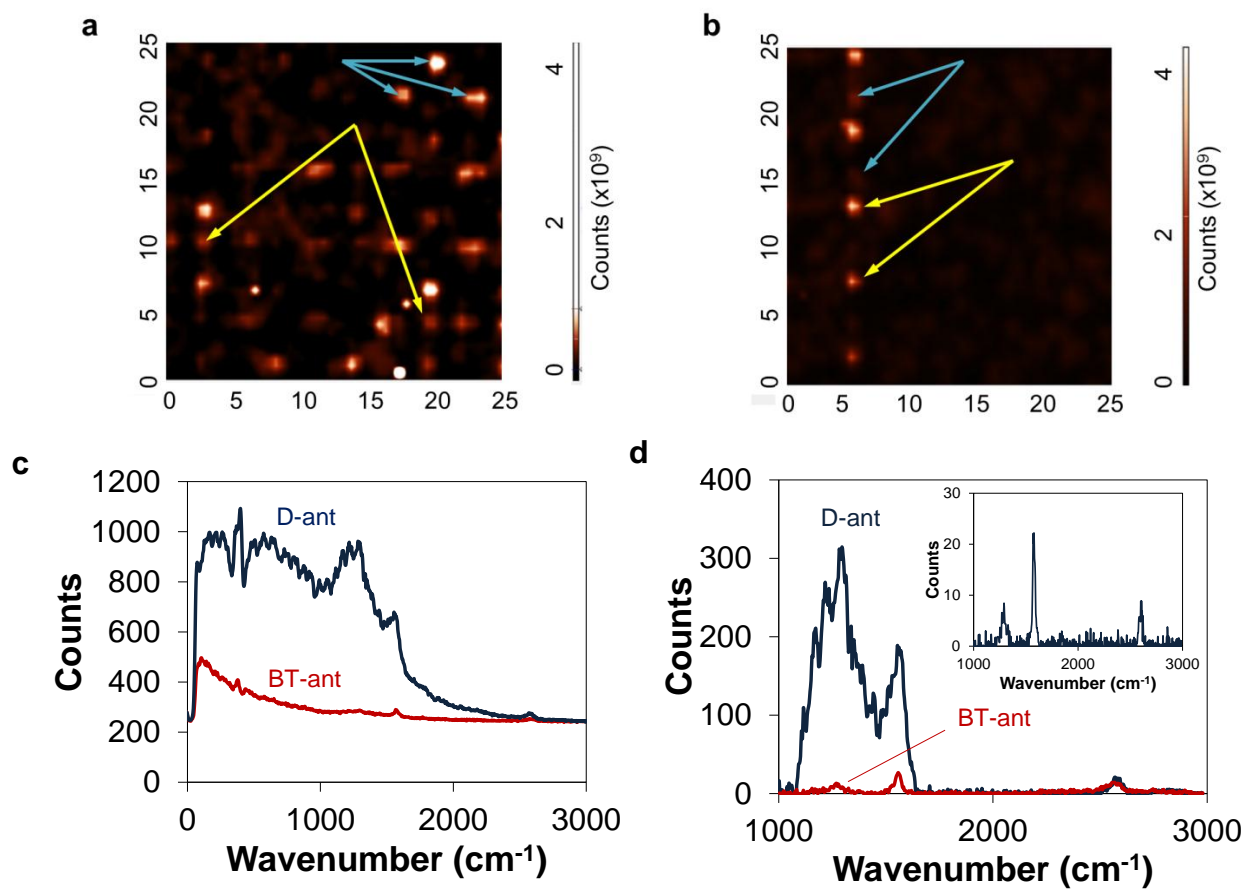

**Figure 5:** Spatial maps and Raman spectra. (a) Uncorrected maps of BT-ant at 2700 cm$^{-1}$. (b) Uncorrected maps of D-ant at 2700 cm$^{-1}$. The laser wavelength was 785 nm and the antenna's gap was nominally 10 nm. The yellow arrows point to the antennas and the light blue arrows point to the marks. An optical filter cut the laser line near zero for both antennas. (c) Uncorrected Raman spectra of both antenna type. (d) Background corrected Raman spectra. The inset shows a typical spectrum of graphene between the D-ant; the relative small amplitude of the 2700 cm$^{-1}$ is due to the limited sensitivity of the Si detector array in this spectral range (~980 nm).

# Raman signal-enhancement and broadening with graphene-coated diamond-shape nano-antennas

*Charilaos Paraskevaidis[1], Tevye Kuykendall[2], Mauro Melli[2], Alexander Weber-Bargioni[2], P. James Schuck[2], Adam Schwartzberg[2], Scott Dhuey[2], Stefano Cabrini[2] and Haim Grebel[1*]*

[1] Electronic Imaging Center, NJIT, Newark, NJ 07102; grebel@njit.edu

[2] The Molecular Foundry, Lawrence Berkeley National Laboratory, Berkeley, CA 94720

Supplemental Information

**SERS signals with 633 nm HeNe laser:** these experiments did not reveal any enhancement whether conducted before or after the experiments with the 785 nm laser. This was true for both antennas' type. Based on these and other spectra, we may conclude that our graphene was made of mostly 2 layers.

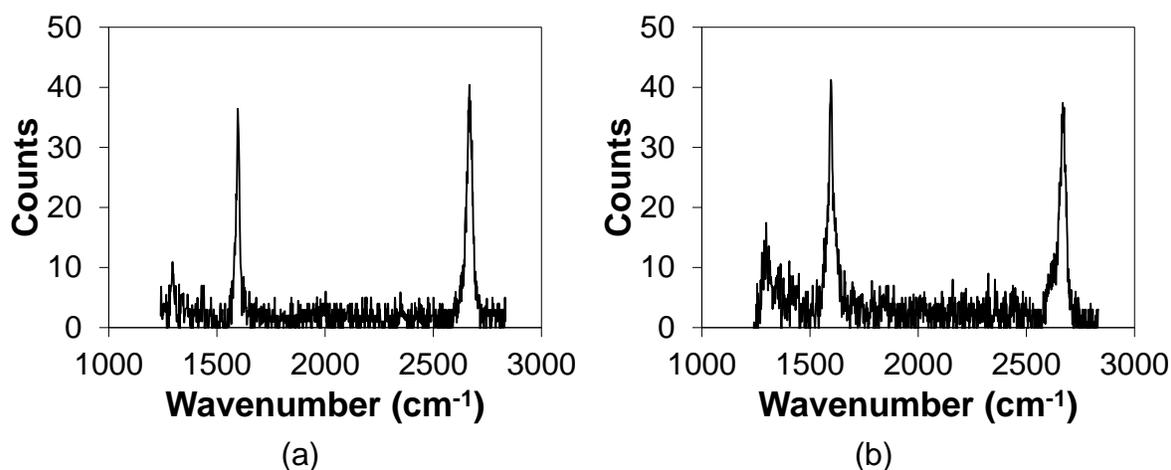

(a)                                (b)

Fig. S1. Background subtracted Raman spectra of graphene coated D-ant taken with a 633 nm laser: (a) the laser was polarized in parallel to the antenna's axis. (b) The laser was polarized perpendicularly to the antenna's axis.

**Model:** The antenna provides gain to all oscillators between the pump and scattered frequencies and the problem becomes nonlinear. It means that the laser light is better absorbed and the Raman signals (and as a matter of fact all other possible oscillators) better radiating out. We assume that the resonance spectral region may approximate Fig. 1c by a line factor, or normalized Lorentzian (namely, a distribution with peak normalized to a constant). The antenna's gain is then written as,

$$G(\varpi) = G_0 \frac{(\sigma_\varpi/2)^2}{[(\varpi^2 - \varpi_0)^2 + (\sigma_\varpi/2)^2]}$$

Here, $G_0$ is the gain amplitude ($G_0 \geq 1$), $\varpi_0$ is the peak resonance wavenumber and $\sigma_\varpi$ is the antenna's spectral width. The large field, associated with the antenna resonance is affecting both the Raman signal and its linewidth. Each of the scattered (Raman) line at $\varpi_i$ may be approximated by a line factor as well,

$$A_i(\varpi) = \frac{c_i \cdot (\Gamma(\varpi)/2)^2}{[(\varpi - \varpi_i + \Delta\varpi)^2 + (\Gamma(\varpi)/2)^2]},$$

where $c_i$ are relative constants of intensity of Raman lines involved. Typically, the related time constants transform as, $1/\tau \rightarrow 1/\tau \cdot (1+I(\varpi)/I_s)$, where $I_s$, is the saturation intensity and $I(\varpi)$ is the scattered intensity at a wavenumber $\varpi$. The time constant is directly related to the line width $\Gamma(\varpi) \sim 1/\tau$. Based on this form we approximate the linewidth by, $\Gamma(\varpi) = [1+G(\varpi)]\Gamma_0$ which implies a very large gain for the scattered mode.

The saturation intensity is defined as $I_s = p/\sigma\tau$ where, $p$ is the emitted photon energy and $\sigma$ is the scattering cross section. If the entire antenna participates in the scattering process, then the cross section is on the order of the antenna's dimension and hence, $\sigma$ is a rather large number. The small line shift might also be attributed to the gain (also known as gain pulling) and is approximated by a small fraction of the spectral width; $\Delta\varpi = a\Gamma(\varpi)$. The spectrum of the Raman signal $I(\varpi)$ is the sum of four active Raman lines $A_i(\varpi)$ at approximately 800 (which could be observed between antennas), 1340, 1600 and 2700 cm$^{-1}$ from the laser line. In terms of wavelengths: the laser wavelength is 785 nm; and the Raman lines are situated at, 834 nm, 873 nm, 895 nm and 983, respectively. By definition, the relative change in the scattered light equals the gain, $\frac{\delta I(\varpi)}{I(\varpi)} = \gamma(\varpi)$. In turn, this gain is related to the amplified incident field through, $\gamma(\varpi) = N\sigma(\varpi)I_L$, where N is the number of graphene atoms involved, $\sigma(\varpi)$ is the Raman scattering cross section peaking at $\varpi_i$. The cross section at the Raman line $\sigma(\varpi_i)$ and the laser intensity $I_L(\varpi_L)$ are correlated. The scattering cross section is inversely proportional to the refractive indices for the laser and scattered mode frequencies [S1]: $[n_i(\varpi_i)n_L(\varpi_L)]^{-1}$. The real value of refractive index in typical Raman scattering experiments is positive ($|n|>1$). However, the antenna resonance imposes, Re{n}→0 on the pump and scattered modes if properly situated within the antenna's gain function. For narrow Raman line widths (otherwise a correlation integral is in order), the scattering cross section at the Raman peak is $\sigma(\varpi_i) = \frac{G(\varpi_i)}{G_0} \cdot \sigma_0(\varpi_i)$ and $I_L(\varpi_L) = G(\varpi_L)I_0(\varpi_L)$, with $\sigma_0$ the non-amplified scattering cross-section and $I_0$ is laser intensity. We generalize to all possible oscillators within the antenna gain curve, $\sigma(\varpi) = \frac{G(\varpi)}{G_0} \cdot \sigma_0(\varpi)$. Each scattered component is, thus weighed by a frequency dependent, antenna gain function. We write for the overall Raman scatterings ($N=\sigma_i=I_L=1$),

$$I(\varpi) = \sum_i A_i(\varpi) \exp[G(\varpi) \cdot G(\varpi_L)/G_0].$$

In Fig. S2 we show fitting to uncompensated data of Fig. 5c. The background increase as $\varpi \to 0$ is a combination of large antenna gain multiplying the Lorentzian tail of the scattered lines. If the Raman lines are very narrow, the antenna simply amplifies the Raman signal with no significant background. When the antenna's bandwidth is narrow and the antenna's peak resonance coincides with the Raman frequency, there would be little enhancement to $I_L$ and, only small enhancement to the Raman signal. In fact, if we hold the model true, part of the background exhibited in our data at ca 800 cm$^{-1}$ may represent an amplified and broaden Raman signal from the tail of the Raman lines.

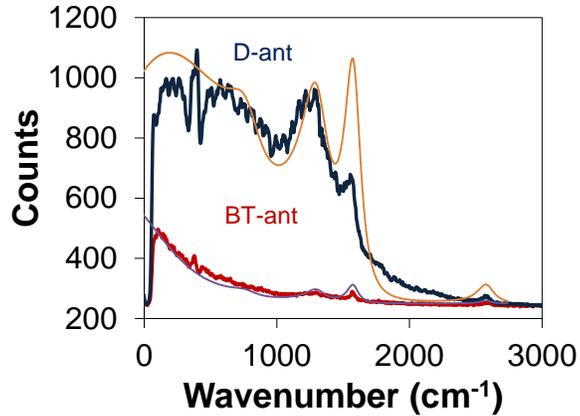

Fig. S2. Fitting Fig. 5c: the laser wavelength was 785 nm. The D-ant's resonance wavelength was taken as $\lambda_0$=780 nm (12820 cm$^{-1}$); the antenna's resonance width was taken as $\sigma_\varpi$=4200 cm$^{-1}$; $\Gamma_1=\Gamma_3=\Gamma_4$=30 cm$^{-1}$; $\Gamma_2$=20 cm$^{-1}$. A constant background of 250 counts was added to the fit. The BT-ant was simulated with the same Raman line parameters, however with peak antenna's resonance wavelength of $\lambda_0$=765 nm (13072 cm$^{-1}$) and resonance width of $\sigma_\varpi$=3100 cm$^{-1}$. The ratio of Raman amplitudes was 0.02:0.3:1:1 for $c_1$:$c_2$:$c_3$:$c_4$, respectively. Other common parameters were: a=0.1, $G_0$=10.8; the graphene density was $N$=1 and the laser intensity was $I_L$=1 cm$^{-2}$.

The fit curves for the two antenna types were obtained with the same gain amplitude $G_0$=10.8, a relative small difference in their respective peak resonance and a 30% bandwidth advantage to the D-ant. We note that the simulated reflection peaks intensities are similar to both antennas indicating that they indeed differ mostly by their related peak resonance wavelength and bandwidths.

**X-Y scans of integrated peaks:** Background corrected plots similarly to Fig. 3 are shown in Fig. S3. Area under the peak profile is plotted here instead of the peak value itself.

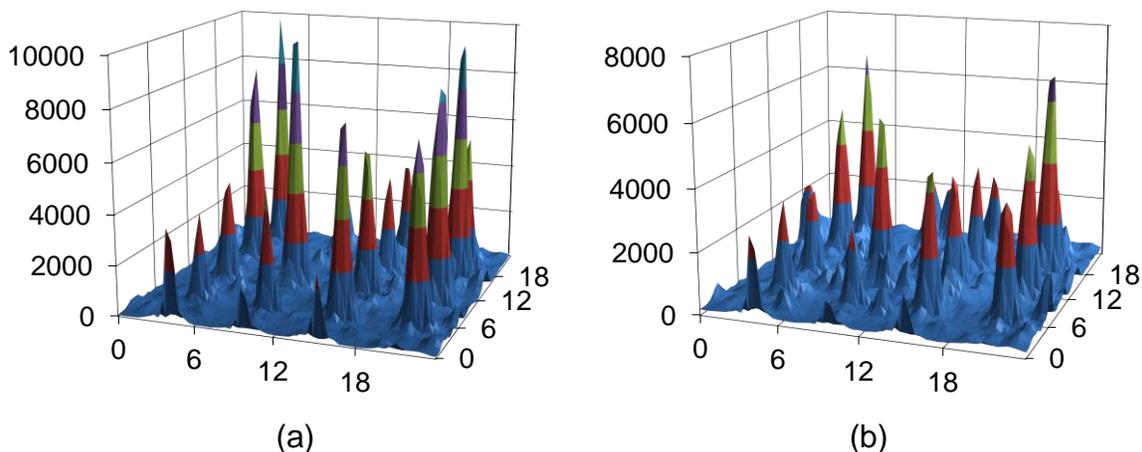

(a)                               (b)

Fig. S3: Background corrected maps: (a) Two-dimensional spatial scan of the integrated 1340 cm$^{-1}$ peak. (b) Two-dimensional spatial scan of the integrated 1600 cm$^{-1}$ peak. The 785 nm laser with parallel polarization to the antennas axis was used.

In Fig. S4 we present data for D-ant, which were excited by a 785 nm laser, polarized perpendicularly to the antennas's axis (after the conclusion of scans with the laser polarized parallel to the antenna axis). Here, the integrated peaks became smaller; the spectral width of the 1340 cm$^{-1}$ and the 1600 cm$^{-1}$ lines (Fig. S4c) remained broad. The fiduciary marks became clearer as well.

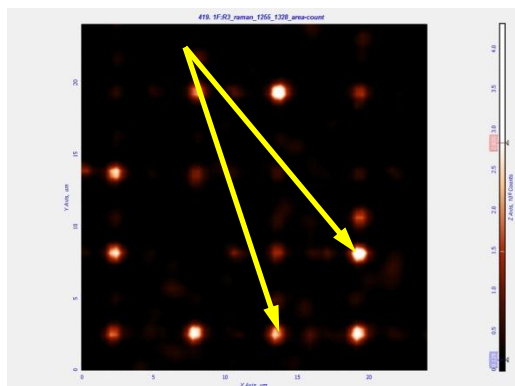

(a)

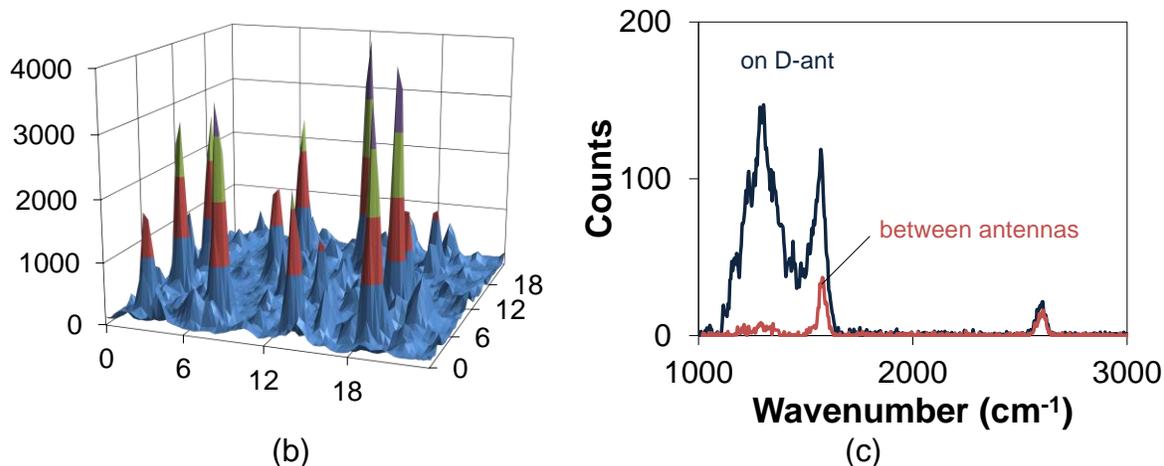

(b)　　　　　　　　　　　　　　(c)

Fig. S4. Using a 785 nm laser, polarized perpendicularly to the antenna's axis after the conclusion of polarized parallel scans.
(a) Uncorrected map of integrated 1340 cm$^{-1}$ nm; the yellow arrows point to the antennas. (b) Corrected map of integrated 1340 cm$^{-1}$ graphene line; (c) Spectrum of graphene on a typical antenna and between antennas.

As a quick assessment we conducted white light scattering experiments of the samples. Obviously, we averaged here over many antennas and marks. Yet, these experiments convey valuable information about the center scattering wavelength. The experiments were carried out with a stabilized and collimated white light source (SPEX), a lock-in amplifier (Stanford Research), a spectrometer (SPEX), and a Si detector. The light was focused onto the sample by means of 10x objective. The focused white light spot, with a diameter estimated at 50 μm, covered quite a few antennas and provided for an average reflection signal. The reflection signal from areas with the antennas was normalized by reflection signal from a graphene–coated area without the antennas. In Fig. S5 we show the experimental configuration and the results for 30 nm gap D-ant and 10 nm gap BT-ant, respectively. The signal from the antennas was referenced to the reflection signal obtained for areas without the antennas. It is clear that the peak reflectance for the BT-ant down shifted to shorter wavelengths, alluding to a larger than expected antenna gap. Upon fitting the white-light data with a Gaussian distribution, we found that the spectral width $\sigma_w$ for the BT-ant was 200 nm - smaller than the 250 nm spectral width for the D-ant and as anticipated by simulations. Yet, the experimental white-light data indicated larger than expected spectral bandwidth and could be attributed to the dispersion of the white light source and inhomogeneous broadening due to the presence of fiduciary marks in the layout.

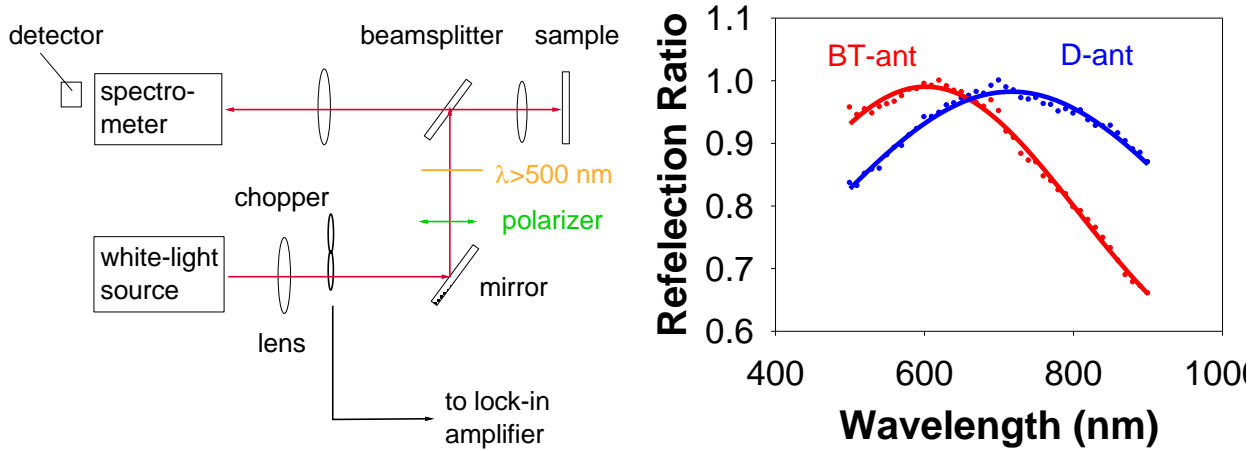

Fig. S5. (a) The configuration for white light experiments. (b) Reflection ratio of 30 nm gap D-ant and 10 nm BT-ant. The reflection signal from the antennas was referenced to the reflection between the antennas. The dots are the experimental data whereas the solid line is a peak fit using a Gaussian distribution. The spectral bandwidth of the D-ant is 0.25 wider than the BT-ant. The incident polarization was parallel to the antenna's axis.

Fig. S6 shows the small but noticeable enhancement and little broadening to the graphene D-line but no amplification or broadening to all other lines when using a 30 nm gap D-ant. There is no down-shift to the Raman lines either.

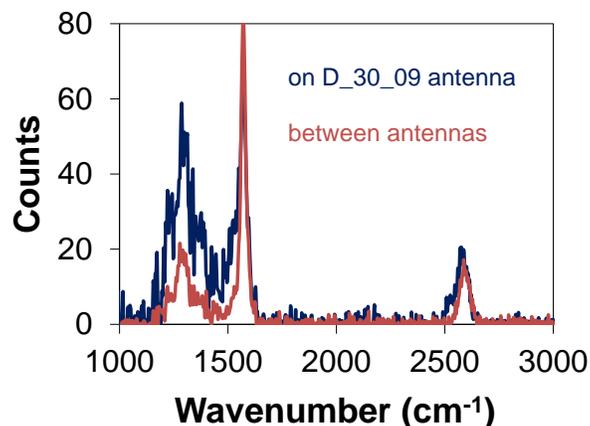

Fig. S6. Raman corrected spectra of a 30-nm gap D-ant and the corresponding reference signal from an area between the antennas. The laser used was 785 nm, polarized parallel to the antennas axis. The small 2D line is attributed to the CCD response.

Finally, in Fig. S7 we show uncompensated data for 10 nm BT-ant and D-ant. The 800 cm$^{-1}$ line is easily identified. The background from 0-500 cm$^{-1}$ is attributed to the oxide substrate. The very broad 800 cm$^{-1}$ line is un-identified when data were collected from the antenna area. Our background correction curves were presented in the region 1000-3000 cm$^{-1}$ where clear lines could be identified.

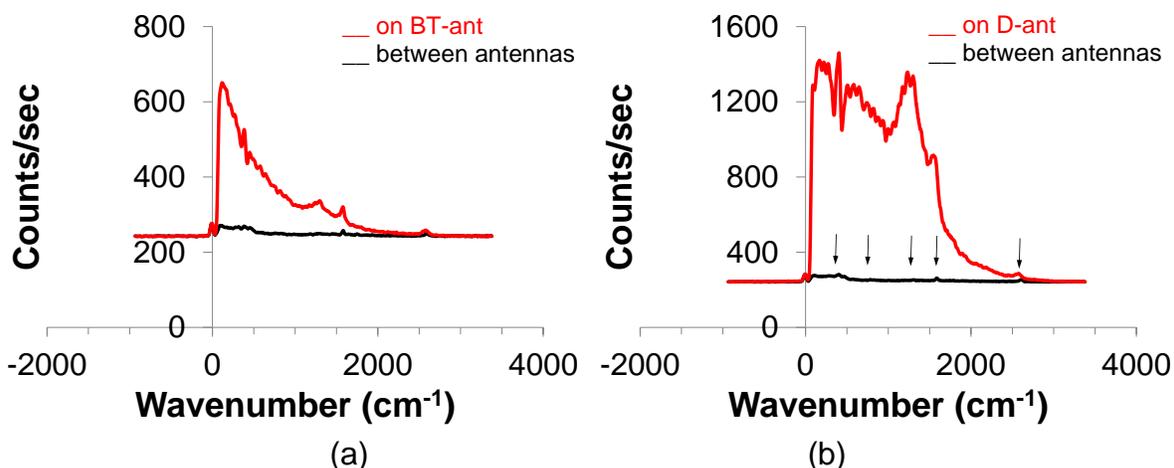

(a)     (b)

Fig. S7. The entire un-corrected Raman data for (a) BT-ant and (b) D-ant. The arrows in (b) identify the lines at 470 cm$^{-1}$ (oxide), 800 cm$^{-1}$ (amorphous carbon), 1300, 1600 and 2700 cm$^{-1}$ of graphene.